\documentclass[aps,prl,twocolumn,groupedaddress]{revtex4-1}

\usepackage{graphicx}
\usepackage{caption}
\usepackage{subcaption}
\begin{document}


\title{On the difference between breakdown and quench voltages of argon plasma and its relation to $4p-4s$ atomic state transitions}


\author{Ebrahim Forati}
\email[]{forati@ieee.org}
\affiliation{University of California San Diego}
\author{Shiva Piltan}
\affiliation{University of California San Diego}
\author{Dan Sievenpiper}
\email[]{dsievenpiper@ucsd.edu}
\affiliation{University of California San Diego}


\date{\today}

\begin{abstract}
Using a relaxation oscillator circuit, breakdown
($V_{\mathrm{BD}}$) and quench ($V_{\mathrm{Q}}$) voltages of a DC discharge microplasma between two needle probes are measured. High resolution modified Paschen curves are obtained for argon  microplasmas including a quench voltage curve  representing the voltage at which the plasma turns
off. It is shown that, for a point to point microgap (e.g. the microgap between two needle probes) which describes many realistic microdevices, neither Paschen's law applies nor field emission is noticeable.  Although normally $V_{\mathrm{BD}}>V_{\mathrm{Q}}$, it is observed that depending on environmental
parameters of argon, such as pressure and the driving circuitry,  plasma can exist in a different state with equal  $V_{\mathrm{BD}}$ and $V_{\mathrm{Q}}$. Using  emission line spectroscopy, it is shown that $V_{\mathrm{BD}}$ and $V_{\mathrm{Q}}$  are equal if the atomic excitation by the electric field dipole moment dominantly leads to one of the argon's metastable states ($4P_{5}$ in our study).

\end{abstract}


\maketitle

Microplasma, which typically refers to a cold (non-thermal) plasma  confined to sub-millimeter dimensions, began to appear in the literature in the mid 1990s.  Besides numerous biological and environmental applications, microplasmas are also being used for gas and surface analysis, and in ultraviolet radiation sources \cite{foest2006microplasmas, papadakis2011microplasmas, li2011role, becker2004non}.  In a DC discharge plasma, the electron impact on gas atoms
or molecules and the secondary electron emission are the two main processes
to sustain the plasma. Such processes are mostly dependent on the
gas characteristics (e.g. pressure and temperature), the electrodes' conditions (e.g. material, surface finish, shape, area, and separation), and the driving circuitry \cite{braithwaite2000introduction, raizer1991gas, yanallah2011numerical, kopeika1975commercial, iwao2004tungsten}. Based on the Townsend's theorem, the breakdown criterion for a DC discharge plasma
in a 1-D geometry can be expressed as 

\begin{equation}
\gamma\left(e^{\alpha d}-1\right)=1,
\end{equation}
in which $\alpha$, also known as Townsend's first coefficient, is
the number of ionizing collisions made by one electron per unit distance
and $\gamma$, also known as  Townsend's second coefficient, is the
number of secondary electrons produced at the cathode per ionizing
collision in the gap. Assuming that electrons move through a uniform
electric field $\left(E=V/d\right)$ and gas pressure $\left(p\right)$,
their mean energies will be proportional to the ratio $V/\left(pd\right)$,
where $V$ and $d$ are voltage and gap size, respectively. The breakdown
voltage is then expected to depend on $\left(pd\right),$ generally
known as Paschen's law  as

\begin{equation}
V_{\mathrm{BD}}=\frac{Bpd}{ln\left(pd\right)+ln\left(A/ln\left(1/\gamma+1\right)\right)},
\end{equation}
where $A$ and $B$ are gas related constants mostly obtained experimentally. The
Townsend's theorem leads to the well-known traditional Paschen curves
for breakdown voltage as a function of $\left(pd\right)$ with a minimum and two upturn arms (approximately V-shaped) \cite{go2010mathematical, leoni2009numerical}. For a fixed gap size on the right arm of the Paschen curve, pressure and $V_{\mathrm{BD}}$ are proportional because of the increase of collisions between electrons and ions/atoms preventing electrons from gaining adequate
kinetic energy for gas ionization. On the left arm
of the Paschen curve, pressure and $V_{\mathrm{BD}}$ are inversely proportional
due to the decrease in the number of species contributing to cascade ionization \cite{tendero2006atmospheric, kopeika1975video, graves1986continuum}. 
\begin{figure}
\includegraphics[width=0.45\textwidth]{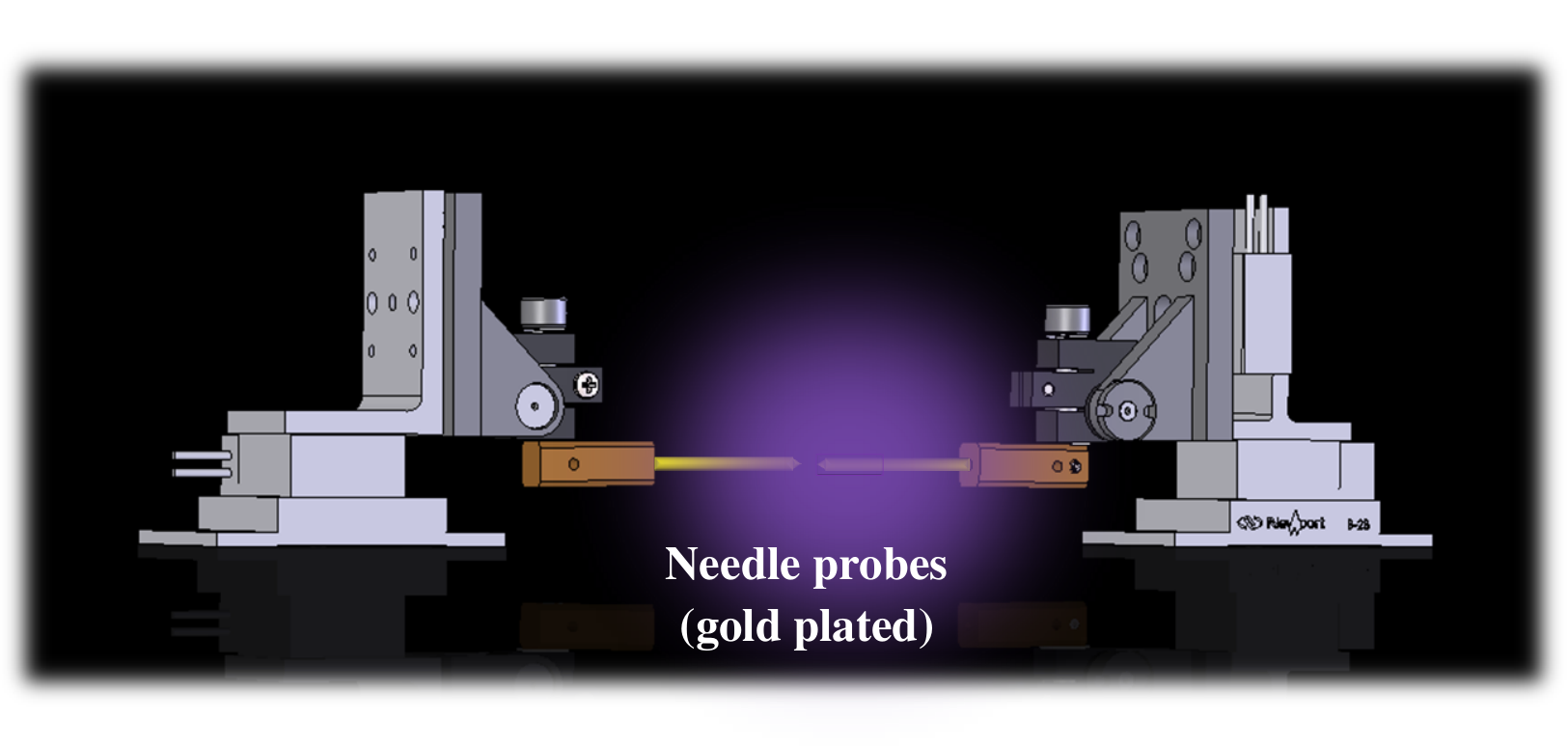}

\caption{The point-point microgap between two needle probes igniting an argon microplasma. The probe holders are mounted on nano-positioners for $xyz$ adjustments and the entire geometry is fitted inside the box chamber shown in Fig. 2.}
\end{figure}

It is suggested by researchers at Bell Laboratories that the electron field emission becomes more significant in small gap sizes at some point. Below $5\,\mu$m in atmospheric pressure, it becomes the dominant process sustaining the plasma. Electron field emission, also called Fowler-Nordheim tunneling, is
the process whereby electrons tunnel through a potential barrier (e.g. a 
metal surface) in the presence of a high electric field \cite{kisliuk1959electron,  radmilovic2013role, radmilovic2008theoretical, spataru1997ion, ecker1959electron, tchertchian2011control, teste1996some} (this is different
from thermionic emission in which electrons emit over the barrier
due to the high temperature \cite{murphy1956thermionic}). The Fowler-Nordheime equation describes the electron
field emission current $\left(I\right)$ as \cite{fang2008inorganic,  kushner2005modelling}

\begin{equation}
I=aE^{2}exp(\frac{-b\Phi^{1.5}}{E}),
\end{equation}
where $a$ and $b$ are constants, $E$ is the electric field and
$\Phi$ is the work function associated with the material of the electrodes.

For microplasmas with small gap sizes, combination
of Townsend and Fowler-Nordheim theorems leads to a modified Paschen
curve, as shown in Ref. \cite{go2010mathematical}. The modified paschen curve replaces the left upturn of the
pure Paschen curve with a plateau and a decline to zero \cite{boyle1955departure, lisovskii2000modified}. 
Nevertheless, both traditional and modified Paschen curves are derived based on 1-D geometries and are confirmed experimentally to be accurate for electrodes much larger than the gap size \cite{klas2012breakdown,  strong2008electrical, radmilovic2012breakdown, radmilovic2005particle, kulsreshath2013ignition, klas2011experimental,   eden2005microcavity, chen2006electrical, 
carazzetti2008experimental}. However, in a practical microplasma device, the size of the electrodes has to be comparable to the gap size to retain all of the device dimensions in the sub-millimeter range \cite{chen2008plasma, sonoiki2013novel}. Along with an ongoing effort to design practical microplasma devices, here we study argon microplasma properties in a point-point geometry, i.e. between two gold plated needle probes with tip diameter $10\, \mu$m and length $5\, $cm, as shown in Fig. 1. The probe holders are mounted on three nano-positioners for $xyz$ adjustments with an accuracy $50\, $nm. 
    
\begin{figure}
\includegraphics[width=0.5\textwidth]{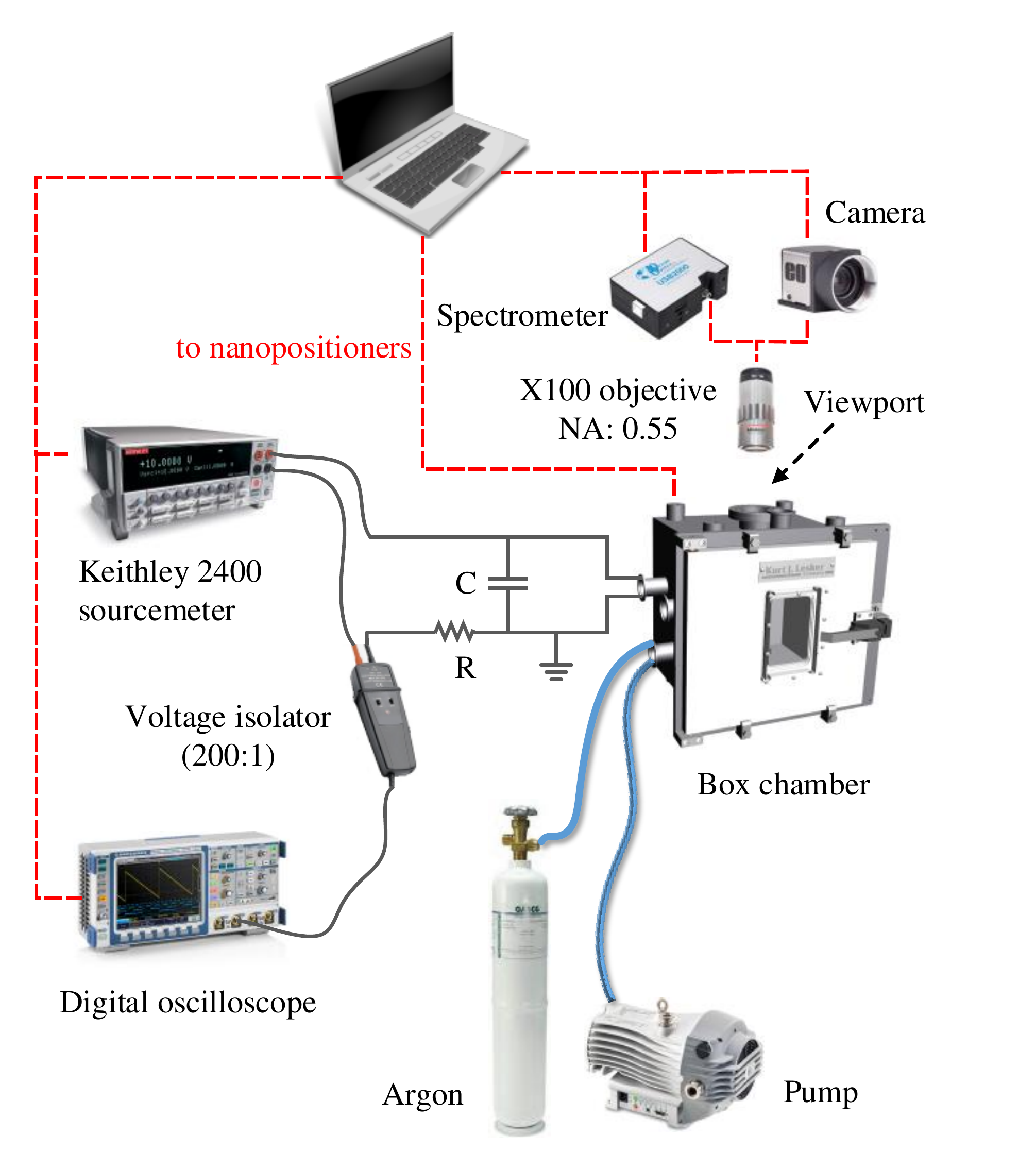}

\protect\caption{The measurement setup.}
\end{figure}

\begin{figure}
\centering
       \begin{subfigure}[b]{0.4\textwidth}
               \includegraphics[width=1\textwidth]{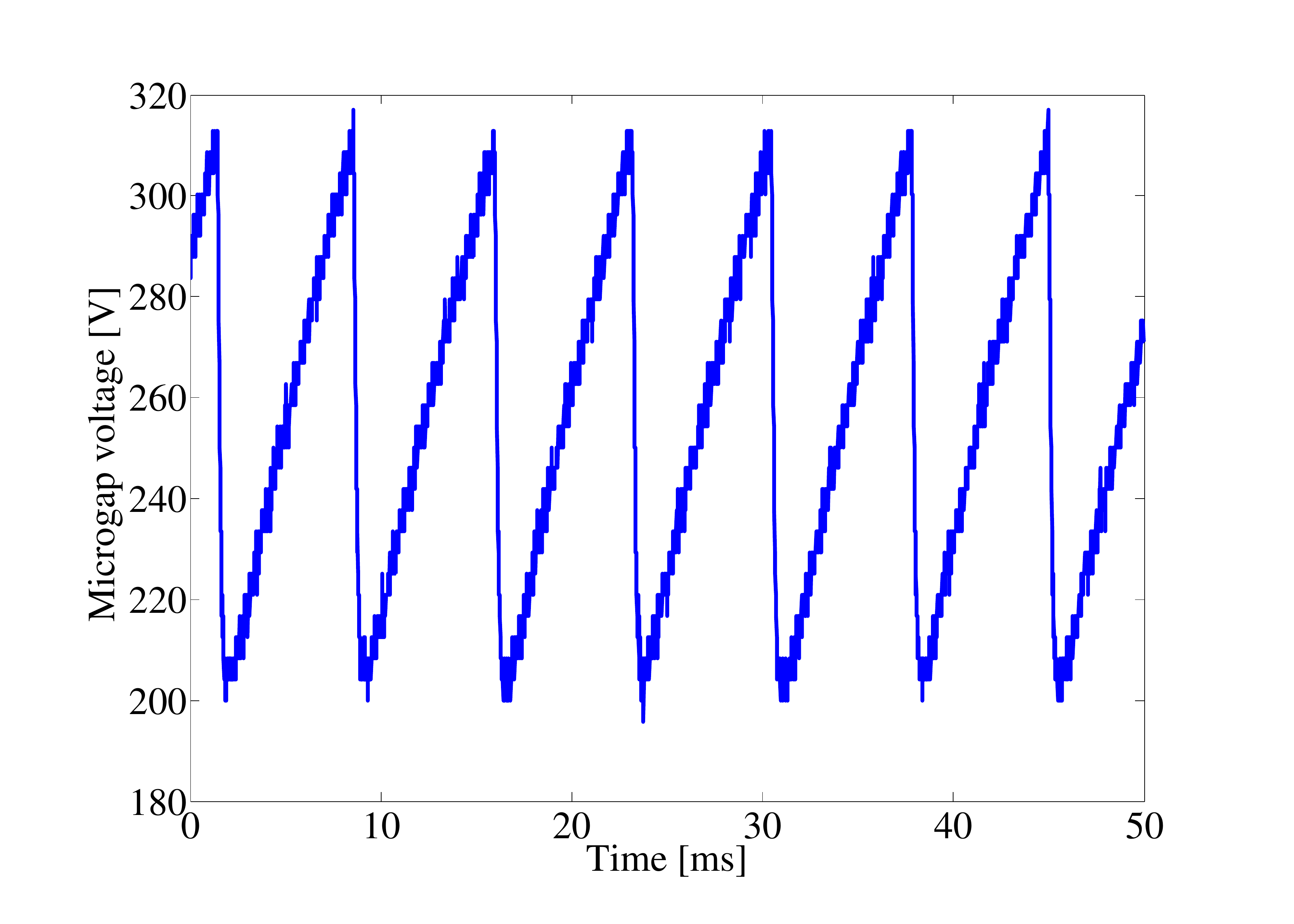}
                \caption{Measured sawtooth across the $500\, $nm microgap at $0.9\, $Torr,}
                \label{fig:3_a}
        \end{subfigure}%
				\vfill
				 \begin{subfigure}[b]{0.4\textwidth}
                \includegraphics[width=1\textwidth]{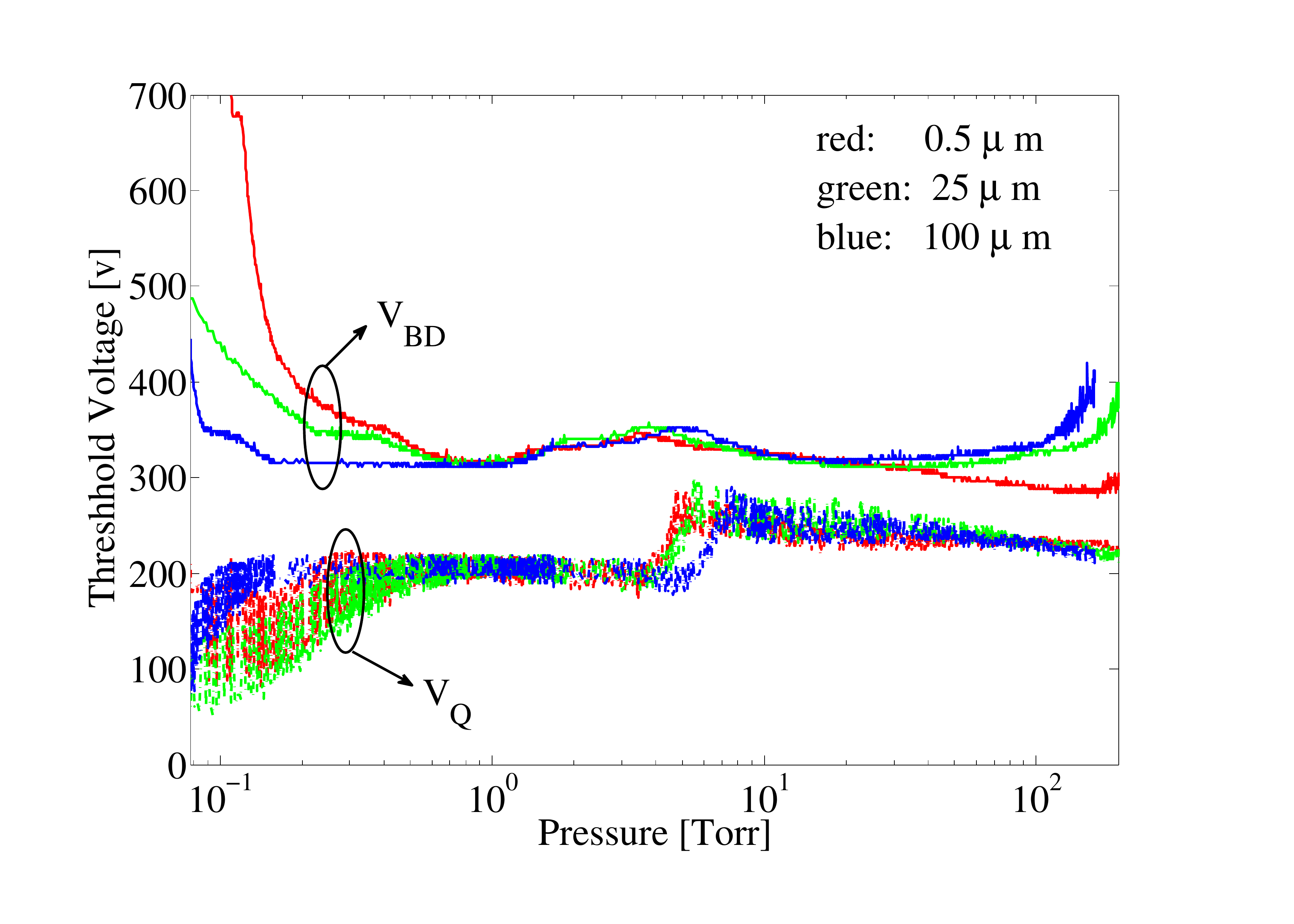}
                \caption{Measured $V_{\mathrm{BD}}$ and $V_{\mathrm{Q}}$ of the argon microplasma with different gap sizes, }
                \label{fig:3_b}
        \end{subfigure}%
				\vfill
				 \begin{subfigure}[b]{0.4\textwidth}
                \includegraphics[width=1\textwidth]{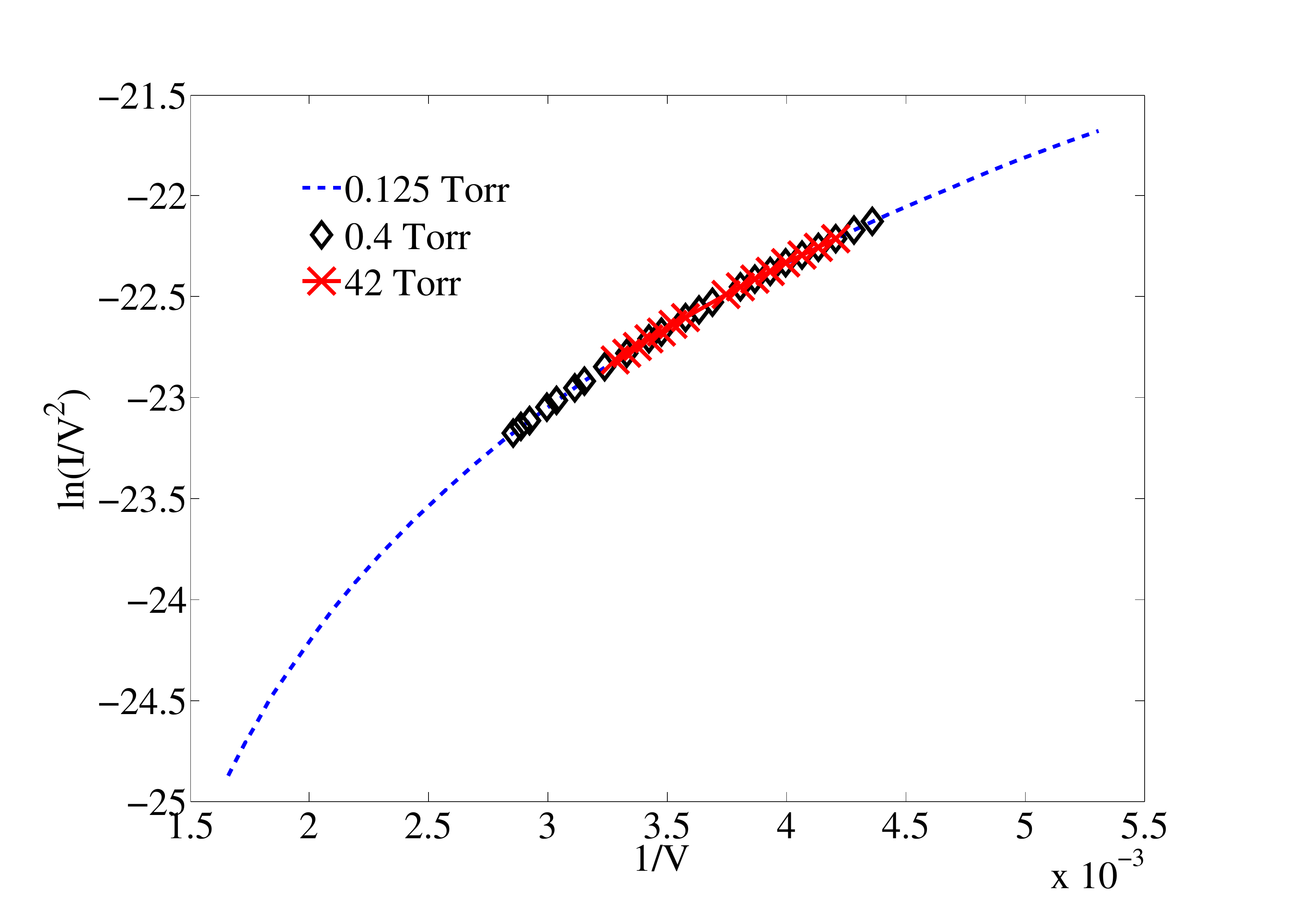}
                \caption{Fowler-Nordheim plot of $500\, $nm gap size at different pressures, }
                \label{fig:3_c}
        \end{subfigure}%
				 \caption{Measured argon microplasma results.}\label{fig:animals}
\end{figure}
Although a DC discharge plasma ignites at $V_{\mathrm{BD}}$, it can normally be sustained
with lower applied voltages since the rise of its free electrons' temperature
and number helps the ionization processes. In other words, usually the
DC breakdown voltage of a plasma is higher than its quench voltage
$\left(V_{\mathrm{Q}}\right)$, in which the plasma turns off, leading to a
hysteresis property for the plasma. Depending on the application, the quench
voltage of a plasma may become a critical parameter as important as
the breakdown voltage (e.g. in switching applications). Despite extensive studies on DC breakdown voltage, we are not aware of any investigations concerning the quench voltage of plasmas. Here, by inserting the point-point microplasma of Fig.1
(with variable pressure and gap size) into a relaxation oscillator
circuit \cite{rca1974relaxation, stoican2009frequency}, we measure its breakdown and quench voltages accurately. 
The measurement setup, including the relaxation oscillator ciruit, is shown  in Fig. 2, in which the sourcemeter provides DC voltage. It
is quite easy to show that due to the difference between $V_{\mathrm{BD}}$
and $V_{\mathrm{Q}}$ of the plasma, a sawtooth voltage appears across the microplasma
with the frequency
\begin{equation}
f=\frac{1}{R_{\mathrm{total}}C}ln\left(\frac{V_{\mathrm{DC}}-V_{\mathrm{BD}}}{V_{\mathrm{DC}}-V_{\mathrm{Q}}}\right)
\end{equation}
in which $V_{\mathrm{DC}}$ is the applied DC voltage  and
$C$ is the parallel capacitor.  The input resistance of the voltage isolator is $2\, $M$\Omega$ and $R_{\mathrm{total}}=R+2 \, $M$\Omega$ is the total series resistance.
As an example, Fig. 3(a) shows the measured sawtooth voltage for $V_{\mathrm{DC}}=700\, $V,
$C=60\, $pF, and $R_{\mathrm{total}}=52\, $M$\Omega.$ The argon microplasma
pressure $\left(p\right)$ and gap size $\left(d\right)$ are $0.9\, $Torr
and $500\,$nm, respectively.

 The extremums of the sawtooth are essentially $V_{\mathrm{BD}}$ and $V_{\mathrm{Q}}$ of the microplasma.
Besides repeatability, this relaxation oscillator technique enables us to extract statistical information about $V_{\mathrm{BD}}$ and $V_{\mathrm{Q}}$ (such as average and standard deviation) simply by reading enough periods
of the sawtooth. Moreover, the current versus voltage $(I-V)$ information
of the microplasma can be obtained easily by sampling enough points
during the breakdown to quench transition (i.e. the sharp transition
in the sawtooth). Using this strategy, it will be shown that field emission is not significant in the point-point geometry
even with small gap sizes.

Figure 3(b) shows $V_{\mathrm{BD}}$ and $V_{\mathrm{Q}}$ of the argon
microplasma as a function of pressure for three different gap sizes including a
very small ($500\, $nm) gap size which does not follow Paschen's law (neither traditional
nor modified versions including electron field emission effect). The relaxation oscillator
parameters used to generate Fig. 3 are $V_{\mathrm{DC}}=900\, V$,  $R=50\, $M$\Omega,$ and 
$C=60\, $pF. The Paschen's law's failure to explain point-point geometries
is a result of the fact that Townsend's theorem is based on a 1-D
geometry (i.e. an infinite plane-plane geometry). Nonetheless, Paschen's
law has been observed \cite{carazzetti2008experimental, chen2006electrical, go2010mathematical} to be an acceptable approximation
for line-line geometries (i.e. the gap between edges of two coplanar
rectangles). In the point-point geometry, different path lengths exist between the two probes where the smallest path length is equal to the gap size. The measured Paschen curve is then the minimum envelope of all possible traditional Paschen curves between the two needles.  On the other hand, the field emission contribution in Fig.
3(b) is negligible because only a small area of each probe (i.e. the tips
of the probe) is exposed to intense electric field. In order to confirm this,
curves of $ln\left(I/V^{2}\right)$ versus $1/V$ of the gap size
$0.5\,\mu $m and at several pressures are extracted from the sawtooth
waveform and shown in Fig. 3(c). This plot, also known as Fowler-Nordheim plot, clarifies if Equation (3) (and therefore field emission) governs the process. The fact that the curves in Fig. 3(c)
are not straight lines with negative slopes, as Equation (3) predicts, proves
that electron field emission is negligible. 

\begin{figure}
\centering
        \begin{subfigure}[b]{0.45\textwidth}
                \includegraphics[width=1\textwidth]{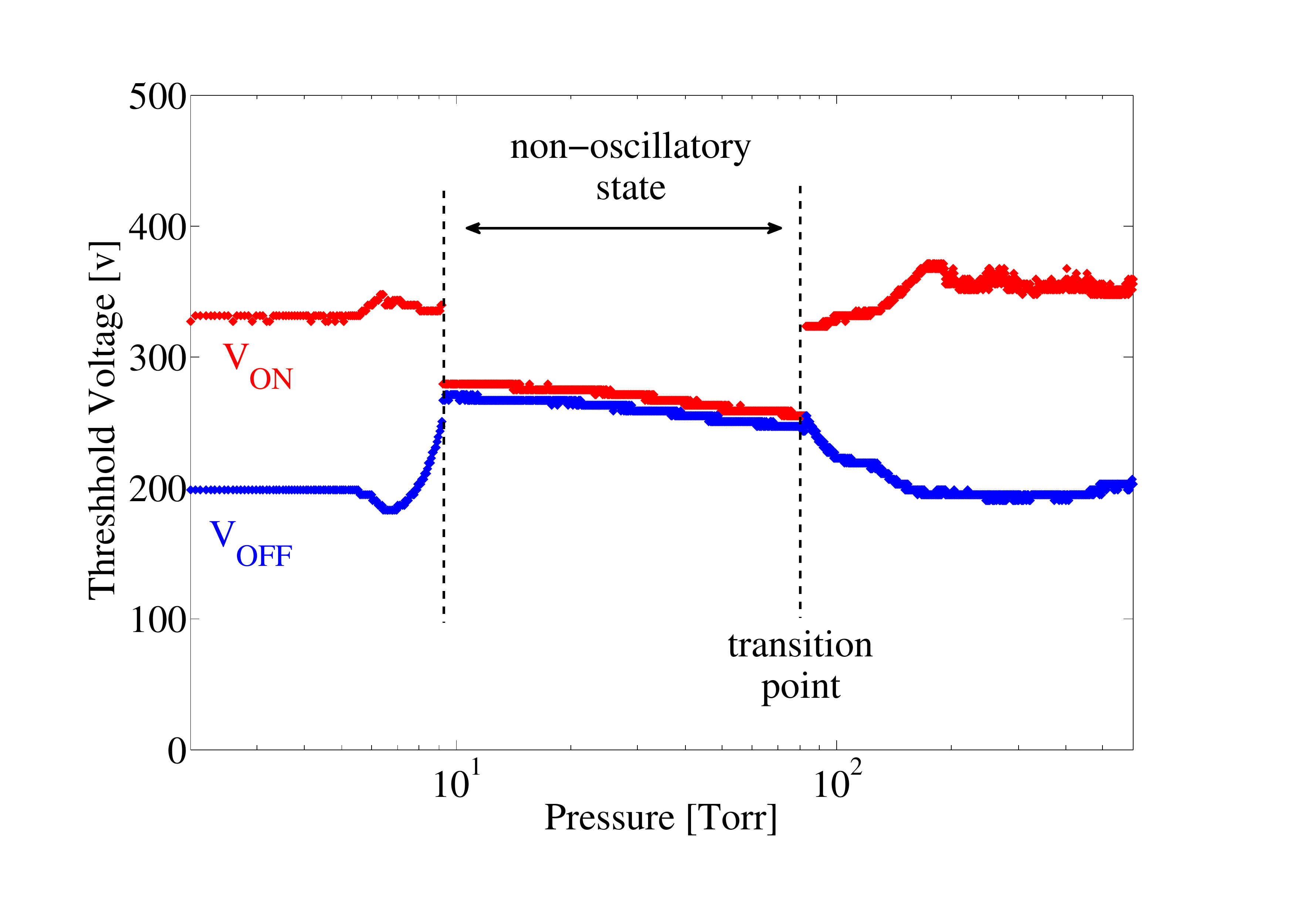}
                \caption{$V_{\mathrm{BD}}$ and $V_{\mathrm{Q}}$ for gap size $64\, \mu $m, $R=2\, $M$\Omega$, $C=60\, $pF, and $V_{\mathrm{DC}}=800\, $V,}
                \label{fig:4_a}
        \end{subfigure}%
				\vfill
				 \begin{subfigure}[b]{0.45\textwidth}
                \includegraphics[width=1\textwidth]{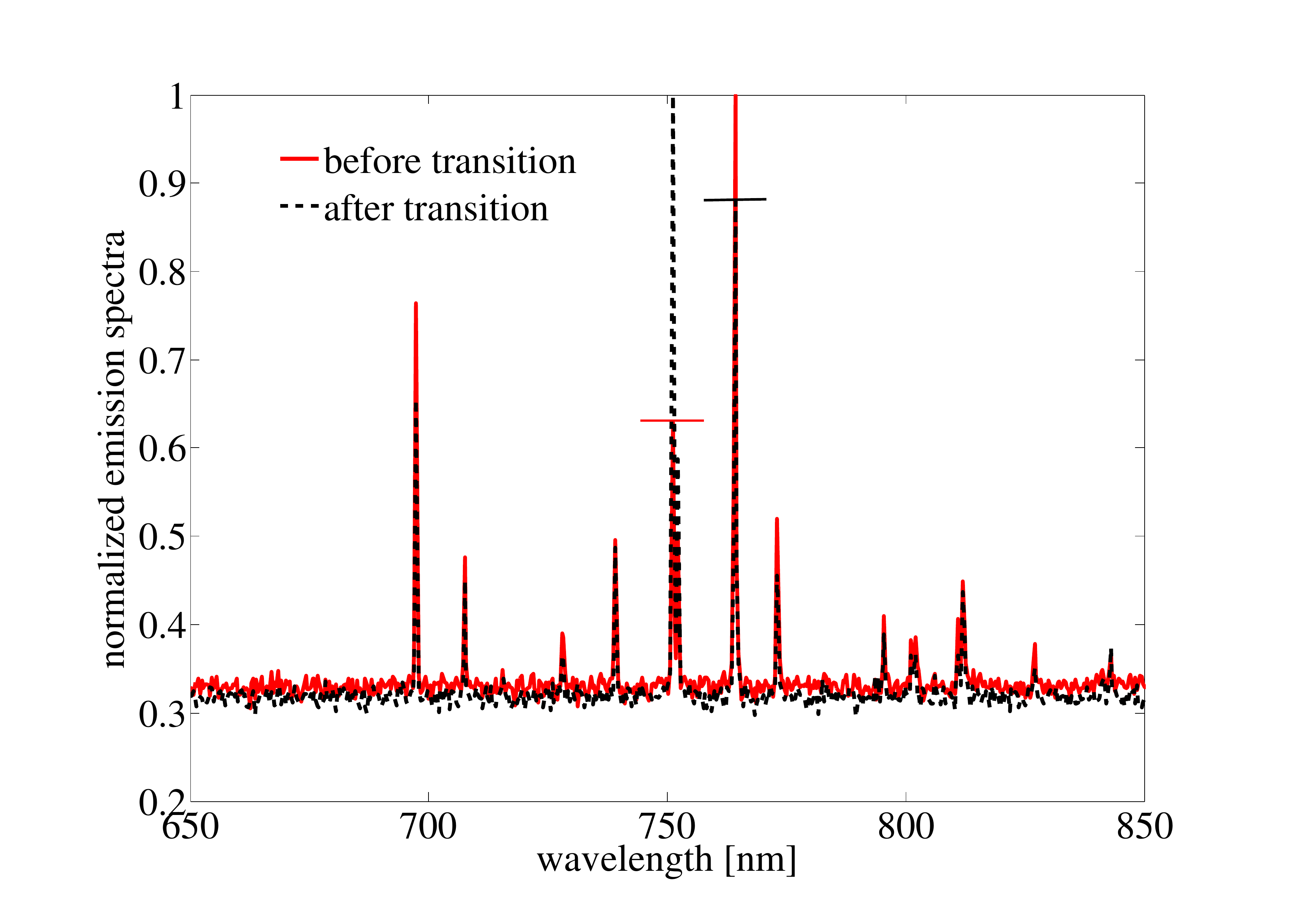}
                \caption{Emission spectra before and after transition from non-oscillatory to oscillatory state,}
                \label{fig:4_b}
        \end{subfigure}%
				\vfill
				 \begin{subfigure}[b]{0.45\textwidth}
                \includegraphics[width=1\textwidth]{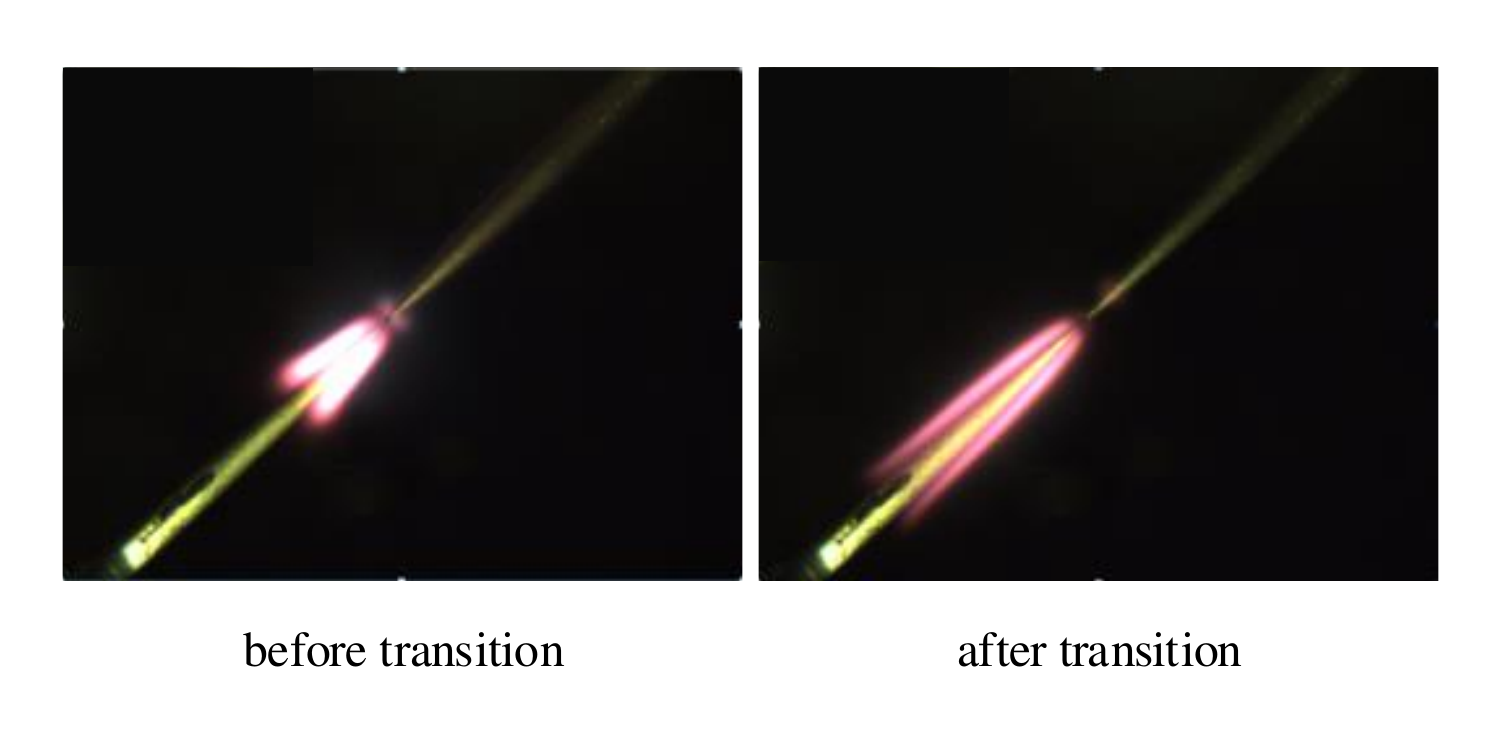}
                \caption{Glow pattern of non-oscillatory (left) and oscillatory (right) states, }
                \label{fig:4_c}
        \end{subfigure}%
				 \caption{Oscillatory and non-oscillatory states of argon plasma.}\label{fig:states}
\end{figure}

It is evident from Fig. 3(b) that both $V_{\mathrm{BD}}$ and $V_{\mathrm{Q}}$ are almost independent of the gap
size and identically depend on the pressure. In fact, several other gap sizes have been measured in the range $0.5-1000\,\mu $m and they all have similar trends. The breakdown voltage has a local peak at $3\, $Torr,
and the quench voltage has minimum difference from the breakdown voltage
at around $5\, $Torr. 

Depending on the plasma parameters such as pressure and the driving
circuitry, electron impact and the electric field dipole moment (by the
applied voltage) can force argon atoms to the excited states with different
probabilities (even zero) of return to the ground state. These different
probabilities relate to the difference between $V_{\mathrm{BD}}$ and $V_{\mathrm{Q}}$.
 There are combinations of plasma parameters in which $V_{\mathrm{BD}}$
can become equal to $V_{\mathrm{Q}}$. For instance, Fig. 4 shows $V_{\mathrm{Q}}$
and $V_{\mathrm{BD}}$ of the point-point microplasma (with gap size $64\:\mu m,$
$V_{\mathrm{DC}}=800\, $V, and $R=0\, $M$\Omega$) with two distinct
plasma states: one with $V_{\mathrm{Q}}\neq V_{\mathrm{BD}}$  is called the
oscillatory state (since it can generate the sawtooth), and the other a non-oscillatory
state with $V_{\mathrm{BD}}=V_{\mathrm{Q}}$ (incapable of producing the sawtooth). Normalized emission spectra \cite{crintea2009plasma, garamoon2007spectroscopic} and glow patterns of the two states are shown in Figs. 4(b) and 4(c) just before and
after the states transition from non-oscillatory to oscillatory behavior as indicated in Fig. 4(a). The argon emission line at $\lambda=763.5nm$,  dominant in the non-oscillatory plasma state, indicates the $1S_{5}-2P_{6}$
argon atomic state transition (in Paschen's notation), and the argon
emission line at $\lambda=751.5nm,$ dominant in the oscillatory
plasma mode, relates to the $1S_{4}-2P_{5}$ argon atomic state's transition \cite{mariotti2006method, miles2010determining, schulze2008robust}.
However, there is a fundamental difference between the final states of
the two transitions (which are $2P_{6}$ and $2P_{5}$ states)
explaining the difference between the two plasma states. The atomic state
$2P_{5}$ is a metastable state, meaning quantum mechanical laws
forbid any electron transition from $2P_{5}$ to the ground state. But, the atomic
state $2P_{6}$ is a resonant state with allowed transition to the
ground state. As Fig. 4(c) shows, most of the electron transitions
in the oscillatory state of the argon plasma lead to a metastable state
where they will be trapped at a high energy level ($11.65\, $eV higher
than the ground state, while the ionization energy of argon is $15.76\, $eV \cite{NIST_ASD}) for a relatively long time (the life time is $56\, $s
 for $2P_{5}$ state). As a result, electrons with lower kinetic energy have the chance to ionize these metastable atoms leading to a lower $V_{\mathrm{Q}}$ than $V_{\mathrm{BD}}$. In other words, as soon as the metastable atoms are generated after the plasma ignition, lower applied voltages (leading to electrons with lower kinetic energy) can sustain the plasma. On the contrary, in the non-oscillatory mode of the plasma, the outer shell electrons of most of the argon atoms experience a step transition
from $1S_{5}$ to the ground state (through $2P_{6}$) leading to lower population of  metastable atoms ( and therefore equal $V_{\mathrm{Q}}$ and $V_{\mathrm{BD}}$).




In summary, we have introduced an easy and practical way to measure both $V_{\mathrm{BD}}$ and $V_{\mathrm{Q}}$ of DC discharge plasmas. Aside from the breakdown voltage, we have studied the quench voltage of microplasmas for the first time. We also have  shown that Townsend and Fowler-Nordheim theorems cannot be used in the form of traditional or modified Paschen curves to predict the breakdown voltage of point-point microplasmas. More importantly, we have shown that depending on the number density of the argon metastable atoms, two different states of plasma, with and without equal $V_{\mathrm{BD}}$ and $V_{\mathrm{Q}}$, can exist.

\bibliography{ref_PRL}

\end{document}